# Reviewing, indicating, and counting books for modern research evaluation systems

**Alesia Zuccala**

Department of Information Studies, University of Copenhagen

**Nicolas Robinson Garcia**

School of Public Policy, Georgia Institute of Technology, United States

## Abstract

In this chapter, we focus on the specialists who have helped to improve the conditions for book assessments in research evaluation exercises, with empirically based data and insights supporting their greater integration. Our review highlights the research carried out by four types of expert communities, referred to as the *monitors*, the *subject classifiers*, the *indexers* and the *indicator constructionists*. Many challenges lie ahead for scholars affiliated with these communities, particularly the latter three. By acknowledging their unique, yet interrelated roles, we show where the greatest potential is for both quantitative and qualitative indicator advancements in book-inclusive evaluation systems.

## 1. Introduction

Since antiquity, books have evolved remarkably. The earliest 'books' were first carved onto clay tablets, and then painted onto papyrus scrolls. In China they were cut into a woodblock, and later, the Europeans printed full manuscripts with ink on paper. Now we have electronic books, or e-books, available online for download. In light of these transformations, seminal volumes have also been written about the history of the printing press (Eisenstein, 1980), book publishing (Thomson, 2005), book classification systems (Mann, 1930), including new perspectives on the book in the digital era (Van der Weel, 2011). There is much to learn from this history in order to review, indicate and count books for modern research evaluation systems.

A valuable starting point is to recognize that books, tightly bound for centuries, are somewhat paradoxical: the information they contained can have power to liberate. Books have been and continue to be change agents in society (Eisenstein, 1980). They change the way that humans think and feel, remind us of our triumphs and follies, and can start a debate or incite a revolution. Some books are lauded; others are not. Some have even been banned from public consumption. Yet, all because of Gutenberg's printing press:

> *a new kind of data collection [was] set in motion even before laboratory facilities were built, or new observational instruments had been invented. The shift from script to print helps to explain why old theories were found wanting and new ones devised even before telescopes, microscopes, and scientific societies appeared. Gutenberg's invention not only preceded Gallileo's tube, it was a more versatile data aid and affected a wider range of data. Some professors shunned controversy and withheld treatises from the press just as some refused at first to look through the telescope. But none failed to consult printed reference guides or preferred to have to copy out tables by hand. Whatever views were held concerning Aristotle, Ptolemy or Galen, whatever objections were posed against using vernaculars or courting publicity; printed maps, charts, and diagrams found rapid acceptance from all* (Eisenstein, 2013, p. 520).

With this level of acceptance, comes great responsibility on the part of evaluators. Books, in essence, capture the efforts of scholars concerned with various types of human endeavors (Garfield, 1979). Yet, for many years the evaluation community has focused on journals articles rather than books. Since the 1960s, the journal article has taken precedence as "a written and published report describing original research results" (Day, 1989, p. 8). In this regard, books and book publishers lag behind, even though they "stand at a crucial crossroad in the production and distribution of knowledge in any society. They are in a position to decide what is 'in' and what is 'out' of the marketplace of ideas" (Coser, 1975, p. 14).

In addition to the journal article, this means that the book needs to be delineated or more clearly defined. In the simplest of terms, Williams et al. (2018) note that "what differentiates a book from a periodical or long report" is that it qualifies for and has an ISBN. Basili & Lanzillo (2018) suggest that an authored book, or monograph, may be defined as "the product of an intense but wide-ranging, systematic and unified research examination of an area of study. Each element contributes to forming the complex of the work, which could not be successfully communicated through the publication of separate parts" (p. 162). The monograph's purpose is to present "what the scholar concludes is the truth about some set of historical events, the characteristics of some work of art or literature or the biography of a historical figure, an artist, or a writer" (Chodorow, 1997). Hence, with a series of scholarly monographs we can piece together the story of a research discipline - i.e., how it has evolved in different regions, over a specific time period, and within a particular interpretive community (Fish, 1980).

In recent decades, the research evaluation community has questioned the value of the book. Adding to this problem has been the decline in sales of scholarly monographs since the 1980s (Thomson, 2005), including a shift on the part of some researchers towards publishing more journal articles (Research Information Network, 2009). Surrounding such publication practices there has also been a lack of stable methods and indicators available to properly assess the monograph's value, impact or influence. Still, research evidence indicates that authored books prevail and continue to prevail because they hold meaning for certain research communities, distinct from those observed in journal literature (Butler & Visser, 2006; Cronin & La Barre, 2004; Cronin, Snyder & Atkins, 1997).

To evaluate scholarly books and account for the influence they have had on their readership, a balancing act is required. On one hand, the evaluative process should respect all that an authored book represents in *qualitative* terms, both to the writer and his/her audience. Book reviews help maintain this respect for quality, since the process of reviewing can be, at the same time descriptive, appreciative, and/or critical. On the other hand, emergent digital tools are now inspiring researchers to devise new ways of assigning symbolic forms of credit to them en masse. Google, Inc., for example, wants all books around the world to 'stand up and be counted' (Taycher, 2010). Clarivate Analytics' decree has been less direct, though critics of the expanding commercial Book Citation Index℠ have much to say about the opportunities and limitations associated with 'putting books back into the library' (Clarivate Analtyics, 2017; Gorraiz et al., 2013)

In this chapter, we review some of the approaches taken thus far to evaluate books both qualitatively and quantitatively. Our focus is on the cluster of information specialists without whom the practice of research evaluation would not prosper: a) *the monitors*, b) *the subject classifiers*, c) *the indexers,* and d) *the indicator constructionists*. This work also provides suggestions for future evaluation systems dedicated to safeguarding book-oriented research fields so that they can continue to develop progressively. As with any guide to evaluation, the "crucial issue at stake is not whether scholars' practices change", but that the application of any specific tool of measurement "enhances research performance and scholarly progress in general" (Moed, 2007, p. 578).

## 2. The Monitors

A monitor may be described as someone who observes, keeps track of, or surveys the progress or quality of something over time. In this sense, many researchers have played monitoring roles, for wider aspects of the research evaluation sys-tems, and for books as well. There is a need for monitors because they show us what is possible to evaluate, where data/information is lacking, and what could be improved upon in the future. While monitors often detect the potential for qualitative or quantitative indicators, they usually do not focus on developing them fully for formal use. They are the historical benefactors of our current system; having brought us to where we are now today with the evaluation of books and book-oriented research fields.

Long before the creation of commercial book indexes– i.e., Clarivate Analytics' Book Citation Index℠ and Elsevier's Scopus – researchers were less interested in metrics for books, and focused more on the uses (or misuses) of published book reviews (Glenn, 1978). For librarians in particular, the book review was and still is considered to be a valuable aid for building book collections (Natowitz & Wheeler Carlo, 1997; Parker, 1989; Serebnick, 1992). Within many library communities it has therefore become essential to study the review culture as a unique form of discourse, and to consider the merits of applying standards for reviewing (Cortada, 1998; Hargrave, 1948; Motta-Roth, 1998; Regnery, 1966). However, with scholars also reading and making use of book reviews (Hartley, 2006; Spink et al., 1998), researchers further recognize that even though a review is not an original work, it can still transfer useful information and ideas. For instance, there is an expectation for a review based on a book published in history to appear in a history journal, but a review could be written about the same book and published also in a political science journal. Lindholm-Romatschuk (1998) explains this transfer in terms of "intradisciplinary and interdisciplinary information flows" (p. 86).

Another critical stage in book-based evaluations took place in the 1970s when researchers began to dissect book reviews using different methods of content analysis (Bilhartz, 1984; Champion & Morris, 1973; Riley & Spreitzer, 1970; Snizek & Fuhrman, 1979). Most of this early work had to be done manually, using datasets of approximately 1,000 to 2,000 reviews. Although researchers today have better technologies for working with data, the first content-based studies initiated a positive trend toward an "informed sociology of the review process" (Snizek & Fuhrman, 1997, p. 114).

The research of Champion and Morris (1973), and also Bilhartz (1984), highlights the degree to which specific time periods have had an effect on review discourse. While book reviews of the 1960s tended to be "gentlemanly" and mostly favorable, those published in the late 1970s and 1980s, specifically for the field of history, increasingly devoted more space to "critiquing rather than simply summarizing a book's content" (Bilhartz, 1984, p. 527). Bilhartz (1984) found specifically that reviewers of the 1970s "took a strong interest in originality of method. However, "more than in any previous decade", reviewers of the 1980s, "expect[ed] histories to have a sharply focused and well-analyzed thesis (pp. 527–528)." Snizek and Fuhrman (1997) consequently hypothesized and found later that favorability in a published book review was significantly and positively correlated to both the age and experience of the reviewer.

Gradually the monitoring phase shifted when information scientists decided to test quantitative techniques for assessing books. Eugene Garfield (1996), creator of the first Science Citation Index, suggested that the creation of a book index would support the *biblio* or book-oriented side of *biblio*-metrics, but, in the absence of this tool, researchers turned to journal indexes in order to analyze book publication, citation, and book review counts (Diodato, 1984; Nederhof, 2006; Nicolaisen, 2002; Schubert, et al., 1984; Thompson, 2002;). Some scholars were working with books as distinct study objects (Lewison, 2001; Hammarfelt, 2011), while others wanted to give more credit to the book as the principal form of publication across the humanities and/or social sciences (Hicks, 2004; Rubio, 1992; Williams, 2009).

With this preliminary stage of book-oriented metrics came the notion that both the humanities and social sciences were at risk of being poorly represented and unfairly assessed (Guillory, 2005). What some of the first *biblio*-metricians did, essentially, was bring to light critical questions about how to assess the social sciences and humanities, primarily because they can be more theory-oriented, and progress more slowly than the sciences (Archambault & Vignola Gagné, 2004). Emphasis was placed on scholars from certain disciplines who might be sharing information using media other than journals (i.e., books!), or contributing to local outlets, including those directed to a non-scholar public (Huang & Chang, 2008; Nederhof, 2006). This led to a significant debate concerning the development and use of alternative databases, like Google Books (Kousha & Thelwall, 2009), or relying more seriously on the open access movement and institutional repositories (Hicks & Wang, 2009; Larivière & Macaluso, 2011; Moed et al., 2009).

Today, the evaluation community can turn to the Book Citation Index℠ and researchers also have the possibility of assessing publication and citation counts to books using the Elsevier Scopus database. But, commercial databases of this nature are a type of 'library', and as researchers subscribe to or become patrons of these unique digital 'libraries', it will be increasingly necessary for them to understand how books are categorized and indexed. This cannot be taken for granted, and in fact with the first Science Citation Index, it was also a primary issue.

## 3. The Subject Classifiers

When Eugene Garfield (1955) first conceived of *a new dimension in the documentation through the association of ideas* (i.e., the Science Citation Index), he reflected on the following:

> *If one considers the book as the macro unit of thought and the periodical article the micro unit of thought, then the citation index in some respects deals in the submicro or molecular unit of thought. It is here that most indexes are inadequate, because the scientist is quite often concerned with a particular idea rather than with a complete concept. "Thought" indexes can be extremely useful if they are properly conceived and developed…. One of the basic difficulties is to build subject indexes that can anticipate the infinite number of possible approaches the scientist may require* (p. 108).

Clearly, subject areas of *thought* were foremost in Garfield's mind, and like the indexes created for journals; subject-based catalogs for books were developed primarily for retrieval purposes. Unlike journals, subject classifications have not yet been used in the development of metric evaluations. However, early monitoring pertaining to book publication and citation counts suggests that a Book Citation Index℠ might indeed be used for this purpose. It is therefore useful to compare subject classes/categories designed for journals with those conceived for books, although a history of the latter is older.

Throughout the 16th to 17th centuries, books held in traditional library stacks were not open for the general public, and available only to special users. Once opened, it became important to position books on the shelves, so that patrons could locate them in relative terms. Melville Dewey, inventor of the *Dewey Decimal Classification System*, recognized that books could be collected together on the basis of similar topics. In 1876, he published the first classification and subject index for books and pamphlets. Several editions of his classification system were published in both English and French (i.e., the French *Classification Decimal*), including an abridged edition, a library edition, and a bibliographic edition, which later became known as the *Universal Decimal Classification*.

After Dewey's death in 1931, the editorship of his classification volumes fell to the Library of Congress. By the time the 16th and 17th editions were published, Dewey's system had been widely adopted by general libraries, but a new Library of Congress Classification (LCC) had also been devised for larger, research-oriented libraries. Both the Dewey Decimal Classification system (DDC) and the Library of Congress Subject Classifications (LCC) possess comparable subject codes and descriptors (see the OCLC Online Computer Library Center, Inc., 2017). The LCC; however, adds an extra Cutter number (a Cutter number refers to the system developed by Charles Ammi Cutter, who invented the *Cutter Expansive Classification system*), which is used to represent a book's author, title, or organization name. While the DDC and LCC are the most predominant systems for classifying books in the United States, other libraries around the world also use them. In countries that do not use a Latin alphabet, alternatives systems have been created, such as the Book Classification of Chinese Libraries (BCCL) in China and the Russian Library-Bibliographical Classification (BBK) in Russia.

Classification systems for journals also support the retrieval of journals and articles based on fields/subjects, but they are also used for evaluation purposes. It is in this realm where classification approaches matter greatly: "reference standards obtained from questionable subject assignment might result in misleading conclusions" (Glänzel & Schubert, 2003, p. 357). According to Glänzel and Schubert (2003), classifications may be produced at different levels of scholarly communication. One may take a cognitive approach, a pragmatic approach, or a scientometric approach. Both the pragmatic and scientometric approaches relate primarily to bibliometric practices, with the first related to journals and the second, individual papers. Commercial journal citation indexes currently use pragmatic subject codes for journals, and normally each subject area is also linked to observed citation patterns. Indexers who monitor these patterns may assign journals to more than one subject category or code (e.g., in Scopus the *New England Quarterly-A Historical Review of New England Life and Letters* belongs to the History ASJC Code 1202 and the Literature Code 1208).

When comparing journals to books, classification systems like the DDC and LCC also produce a code, each approximately 6 to 10 digits in length. For example, the LCC number for the book titled *Uncensored War: The Media and Vietnam* is: DS559.46.H35 1986. Here, the pragmatic approach to classification is also retrieval-based, but it is subject further to *literary warrant*: an LCC can only be produced on the basis of what the classified literature and controlled vocabulary of that time warrant (Beghtol, 1986; Chan, 1999). The first two lines, separated by a decimal, refer to the subject of the book. The third line represents the name of the author, and the last line is the book's publication date. When the library patron finds this call number in a catalog, he or she can go to a section of the library and locate the exact book. Replicas of the same book may be on the shelf, including others related to the same topic, but the book does not appear in two different shelf locations (e.g., the history and the political science shelving area), even if it contains information pertaining to both subjects. In sum, books differ from journals because normally they are fixed to one subject class or category.

Fast-forward to the digital age and the new Book Citation Index[SM] and it is still unclear what Clarivate Analytics' means by *putting books back into the library*? How will this new digital *library* contribute to an evaluation context? More specifically, how can traditional subject classification systems for books, like the DDC and LCC support *metric* evaluations? At present, none of the traditional book classification schemes have been incorporated into the Book Citation Index[SM]. Instead, one finds categories and keywords, which have yet to be fully explained. For example, one can look for the book *Epicureans and atheists in*

*France, 1650-1729* (Kors, 2016) in a traditional library catalog and the classification will be either an LCC B573.K67 2016 or DDC 194–dc23 (see http://lccn.loc.gov/2016008144). In the Book Citation Index℠ this same title is simply classified under the following key terms: *History; Philosophy; Religion*.

The level of granularity afforded by classifications like the LCC and DDC is thus overlooked and may be problematic for book metrics, given what we know for journals. The classification of journals by field/subject is considered to be "one of the basic preconditions of valid scientometric analyses" (Glänzel & Schubert, 2003, p. 357). Journal categories are used, for example, to map the structure of science (e.g., Boyack et al., 2005; Leydesdorf & Rafols, 2009), to normalize impact factor values, and to aid in the calculation of impact factor windows (Dorta González & Dorta González, 2013; Dorta González & Dorta González, 2015; Glänzel & Moed, 2002). So far, little research has been done to reflect the role of subject classifications as a precondition for book or *biblio*-metrics (Chi et al., 2015; Glänzel et al. 2016; Zuccala & White, 2015). This is in part due to the current structure of the Book Citation Index℠. A solution is needed, particularly for the social sciences and humanities, since these fields are more strongly represented in this index than in any other databases of the Web of Science (Chi et al., 2015; Gorraiz et al., 2013).

To circumvent the classification problem, at least two approaches to have been taken. The first, devised by Glänzel et al. (2016), was to match the current Web of Science classification scheme to 74 subfields from the modified Leuven-Budapest classification scheme. With this combined classification approach, clear differences were found for the citation impacts of humanities books versus those of journals in the same field. Another method has been to use an Application Programming Interface (API) to match titles of cited books, retrieved from Scopus to the same titles recorded in OCLC-WorldCat union catalog (e.g., Zuccala & Cornacchia, 2016; Zuccala & White, 2015). Following this matching process, Zuccala and White (2015) were thus able to classify a selection of titles from Scopus history journals (published in 1996 to 2000 and 2007 to 2011) according to their respective DDC classes.

## 4. The Indexers

While a subject classification system is essential to both the retrieval and evaluation of books, a metadata framework designed to catalog them is also needed. We separate the indexers from the subject classifiers because the decisions that these specialists make with regards to metadata also affect the practice of *biblio*-metrics, but in a different way. In short, the data that can be analyzed is only as good as how accurately it has been indexed; hence the process of indexing thousands of books in The Book Citation Index℠ has become both a research topic and subject for serious scrutiny (Gorraiz et al., 2013; Leydesdorff & Felt, 2012; Zuccala et al., 2018).

Not long after the Book Citation Index℠ was launched, Gorraiz et al. (2013) produced some test analyses and found that "out of the almost 30,000 'books' retrieved [for] the publication period 2005–2011, only about 1,100 provide[d] author affiliations" (p. 1390-1392). In addition to missing address information, the researchers noticed that the term 'book' as a registered document type had potential to be confusing, especially if edited books were not carefully delineated from the 'whole book content' of monographs. In cases where there was no clear delineation, there was further potential for false-interpretations, for example: "selecting 'books' as document type and sorting the results by most cited…present[ed] a list of the most cited books as a whole, but disregard[ed] all the citations to single chapters…. Similarly, sorting 'book chapters' by times cited omit[ed] whole-book citations" (p. 1392). Index-focused research also suggests that there can be a problem of underrating or overrating the citation impact monographs if individual chapters from specific monographs are counted separately (Leydesdorff & Felt, 2012).

In comparison to journal articles, monographs are difficult to index because they typically belong to bibliographic families (Zuccala et al., 2018). Unlike journal articles, they can be revised and reprinted as new editions. In the past, many book catalogs have benefitted from guidelines like Function Requirements for Bibliographic Records (FRBR) (Tillet, 2001; 2005); hence Zuccala et al. (2018) suggest that the Book Citation Index℠ can as well. With Tillet's (2001; 2005) conception of the FRBR model, every monograph in a bibliographic family is a physical entity or *manifestation* with its own International Standard Book Number (ISBN). If several different *manifestations* share the same intellectual properties, they are *expressions* (editions), and together all derivative *expressions* (editions) relate to one *work* (see Zuccala et al., 2018). A *work* is therefore the progenitor for a bibliographic family; the starting point for all ideational and semantic content (Smiraglia, 2001). Any new *expression*, or edition of a monograph that deviates significantly from the progenitor is called a new *work*.

The Book Citation Index℠ might potentially be revised to follow FRBR, so that every *expression*, or edition of a monograph is indexed according to its full set of manifestations (i.e., all ISBNs per physical type), its own unique *expression* identifier, and its shared *work* identifier. For each manifestation of a particular book, there is; however, a specific problem to consider. Books, unlike journal articles, do not have their own unique Document Object Indicators (DOIs). Currently the International Standard Book Number (ISBN) is and has become the most frequently used identifier for retrieving and matching identical book titles recorded in different databases (Kousha et al., 2017; Zuccala & Cornacchia, 2016). It is im-portant; however, to recognize that: a) ISBNs do need to be registered, as this gives a clear idea of how many times a book as been re-printed, and b) publication and citation counts should not be calculated at the level of the ISBN, as this does not correspond with the intellectual content of a work, but its physical container. With a proposed FRBR-guided version of The Book Citation Index℠, all ISBNs per book would be present, but the addition of new identifiers means that bibliometricians might have more accurate options for counting books either at the *expression* level, or the *work* level. Zuccala et al., (2018) explain why this matters:

> *The value in calculating indicators at different bibliographic levels is that it can help to identify whether or not a specific expression or edition of a monograph is receiving more attention than the work as a whole. For instance, one specific expression of a work may be cataloged in libraries, used, referred to, or reviewed more frequently than another. This could be the literal translation of a non-English edition of a work to English, with the new English-language edition potentially having a wider appeal. For some types of translated works, in fact, an author might even have more than one metric profile.* (p. 156)

## 5. The Indicator Constructionists

Indicator constructionists are researchers who develop indicators for use in quantitative research evaluation systems. This group of experts differs from the monitors because they are less intent on describing approaches to *biblio*-metrics and more committed to identifying and promoting real methodological solutions. Progress in this regard has been aided greatly by technological advancements and the emergence of new data sources, for example, the Book Citation Index℠, the Scopus index of books, Google Books, Google Scholar, OCLC-WorldCat, Goodreads, Amazon Reviews, and national academic repositories (e.g., Giménez-Toledo, 2016; Kousha et al., 2016; Zuccala & White, 2015).

The process of evaluating books depends, however, on more than just data. When a particular data source is used to advance an indicator, advocates of that indicator need to reflect to some degree on a theory (Gingras, 2014; Zuccala, 2016). According to Zuccala (2016) the main task of the humanistic *biblio*-metrician, or book evaluation specialist is not to simply "expand his/her metric toolkit, but to first examine the term *indicator*" (Zuccala, 2016, p. 159). Gingras (2014) upholds this notion by explaining that if an indicator serves as a proxy of a concept, it must be closely aligned with the concept or object that it is designed to measure. The primary, ongoing difficulty is that "the reality behind the concept [might] change over time and/or place" (p. 113). In Van der Weel's (2011) "*Changing out textual minds*" we are reminded of this fact for books:

> *digitisation of textual transmission is proceeding so rapidly that already the consequences are huge and all-encompassing, indeed revolutionary. As reading practices move on line the once discrete products of the print world all become part of the digital textual 'docuverse', and that docuverse in turn becomes part of the all-digital array of mediums converged on the WorldWide Web* (p. 2).

Will bibliometric evaluations manage to keep up with this revolution?

In terms of data and theory, the research community thus far has taken two paths towards developing book indicators. One route has been to focus on the traditional citation - e.g., extracting citations to books as non-sourced items in commercial indexes (Butler & Visser, 2006; Hammarfelt, 2011; Linmans, 2010). Another has been to avoid the citation and focus on book reviews (Kousha & Thelwall, 2016a; Zuccala & van Leeuwen, 2011), publisher quality and specialization (Torres-Salinas et al., 2014; Zuccala et al., 2014a) and library holding counts (Torres-Salinas & Moed, 2009; White et al., 2009).

*Citations*

In principle, a new data source like the Book Citation Index℠ could seem like the perfect solution for developing indicators for books. Still, there are certain factors to take into account. Research has shown that citation patterns for books differ from that of journal articles (Cronin et al., 1997; Torres-Salinas et al., 2014a) and that in comparison to journal articles, the citation age for books is longer (Glänzel et al., 2016). The role that a book plays within a particular scholarly communication system also differs depending on the discipline under study (Kousha & Thelwall, 2009). And, even within different disciplines there can be citation effects related to book types (Milojević et al., 2014; Thelwall & Sud, 2014), language and internationalization (Engels et al., 2012; Verleysen & Engels, 2014) and variations in authorship patterns (Ossenblok et al., 2014).

With the Book Citation Index℠, the drawbacks to developing new indicators rest with the selection bias of monographs published in the English language, a high concentration of books printed by big publishers, and unclear distinctions between different editions and translations of the same monograph (Torres-Salinas, 2014; Zuccala et al., 2018). There is; however, at least one benefit to this index in that it enables large-scale comparative analyses of citation distributions for both monographs and journal articles (Chi, 2016; Glänzel et al., 2016).

In 2004, Google launched two revolutionary services: Google Books and Google Scholar. Both services not only offer quick and easy access to scientific literature, but also give researchers an opportunity to engage in full-text searching. This in turn enhances the ability to capture citations from a great variety of research sources. The downside to these platforms is that mechanisms by which researchers can identify citations often bring false positives and prevent opportunities for large-scale analyses (Kousha & Thelwall, 2009). Thus, when using Google Books or Google Scholar, researchers suggest that it may be wise only to use citation data as a complement to peer review (Kousha et al., 2011).

Citations have been used outside the scholarly communication system to assess the non-scientific impact of research where scholarship may be targeting a non-scholarly public, intentionally or not. For instance, citations from Wikipedia, which are now part of the set of indicators offered by the platform Altmetric.com, have been suggested as a means to capture extra evidence of impact (Kousha & Thelwall, 2017). Scarce counts; however, makes the 'Wiki-cite' unreliable for use in a real research assessment exercise. There are also many syllabi and teaching materials that include citations to research, which means that books may be measured further in terms of their educational impact (Kousha & Thelwall, 2016b). Since correlations between educational-based citations and research-based citations tend to be low, educational impact is arguably a different type of measure, warranting further investigation on its own.

*Publisher prestige or quality*

With the study of books, an analogy may be drawn between journals and publishers. Though, unlike measures for journals - i.e., the Journal Impact Factor (JIF) (Garfield, 2006), the Source Normalized Impact Per Paper (SNIP) (Moed, 2010), and the SJR (González-Pereira, 2010) - there is currently no similar impact-based quantitative indicator for books. The main focus; therefore has been to assess publisher prestige or quality, instead of impact, and to direct this towards expert (scholars') opinions rather than citations (Giménez-Toledo & Román-Román, 2009; Giménez-Toledo et al., 2012). Proponents of this research area argue that citation data does not accurately capture the impact of books (Giménez-Toledo et al. 2013), and that this is particularly the case in in many humanities disciplines, where the goal is not to create impact per se, but to influence further academic thinking and/or debate (Zuccala, 2012). The *expert-oriented* approach is, or has largely been inspired by the work of Nederhof et al. (2001) who first studied publisher quality within the field of linguistics. In this study, scholars from The Netherlands, Flanders and worldwide were invited to participate in a survey. With the results, Nederhof et al. (2001) were able to differentiate amongst the three different populations and obtain insights into the locality of prestige, language biases and disciplinary differences: all issues which are considered to be highly relevant within the social sciences and humanities (Hicks, 2004).

As a result, we have seen at least one indicator that has been developed and proposed for the evaluation of book publisher 'quality' and 'prestige'. In the research by Giménez-Toledo et al. (2012) 14 questions were sent to various academic/scientists from different research fields as part of a survey that was "structured in three blocks: (1) Profile of the respondent; (2) Evaluation of the quality of a publisher with scientific publications; and (3) Evaluation of the publishing process of a publishing house with scientific publications" (p. 67). Following the survey, the data were used to calculate what the authors' term, an "*Indicator of Quality for Publishers according to Experts (ICEE)*" (Giménez-Toledo et al., 2012, p. 68).

Not all scholars agree with the focus on publishers in the development of book metrics. Verleysen and Engels (2013) indicate, for instance, that publishing houses are not the most suitable level of aggregation, and argue that it is impractical to perform a 'quality analysis' for each and every book title, after it has been published. As a compromise, they suggest creating a label for peer-reviewed monographs so as to ensure that researchers and evaluators know that a certain level of formal quality has been ascertained prior to publication. In this way, more emphasis is placed on the pre-condition for book quality rather than a metric analysis of quality *ex post facto*; a point which continues to be under international discussion (Giménez-Toledo & Román-Román, 2009).

Yet another area of interest has been the study of publisher specialization (Giménez-Toledo et al., 2015; Mannana-Rodriguez & Giménez-Toledo, Torres-Salinas et al., 2012). To understand specialization, one approach has been to take a specific unit of analysis, like the book chapter and develop mapping techniques designed to visualize their disciplinary profiles (Torres-Salinas et al., 2013). Network maps, which follow directed citations to books from journals have also been used, both to identify the specialization of commercial as well as university presses (Zuccala et al., 2014a). In the research of Mannana-Rodriguez & Gimenez-Toledo (2015) the tension between publisher specialization and multidisciplinarity has been measured using what the authors call an "entropy-based indicator" (p. 18). When a publisher publishes books in fewer fields, its specialization increases, whereas its multi-disciplinary profile may also increase if there is unevenness in its distribution of titles across different fields.

To date, only a few publisher rankings have been produced, and only for certain research fields (e.g., SENSE, 2009; Zuccala et al., 2014a). Within a specific time frame, a ranking of publishers may be calculated on the basis of their overall received citations or average citations per book (Zuccala et al., 2014a). However, if a ranking is based on citations, usually the most powerful English-language publishing houses are listed. This is because a large majority of publishing houses tend to have high rates of un-citedness (Torres-Salinas et al., 2014b). Citations only reveal a small portion of what is happening in the publishing industry. A careful ranking procedure must therefore consider the fact that every publishing house or press differs in terms of economic capital, symbolic capital, and geographical reach (Thomson, 2005; Zuccala et al., 2014a).

*Book reviews*

According to Lindholm-Romantschuk (1998), the difficulty attached to finding an appropriate quantitative indicator for assessing book quality, is that processes of formal assessment are and already have been taking place. For a long time "the evaluation of scholarly monographs [has been] contained within the system of academic reviewing" (p. 36). Book reviews still play an important role in the reception of scholarly monographs, but the lack of esteem attached to reviews has sometimes led to legitimate concerns regarding their judicious value (Glenn, 1978). It can be useful therefore to filter out specific types of reviews, by focusing on those that are more 'scholarly' or at least those that researchers agree upon as having familiar or trusted scholarly characteristics (Hartley, 2006; Nicolaisen, 2002). Evidence of scholarliness can be assessed, for example, by the degree to which a reviewer includes references to other academic sources in addition to the book under review (Zuccala & van Leeuwen, 2011).

Yet another way to import book reviews into an evaluation context is to make use of them as *mega-citations*. Zuccala et al., (2014b) have introduced a theory of 'mega-citation', which explains how book reviews may be transformed into quantitative indicators based on a full-text analysis of reviewer comments. One drawback to working with 'mega-citations' is that full-text reviews published in journals are largely inaccessible in electronic form. In light of this problem, some bibliometricians have found that public and socially motivated book reviews are a better option, particularly those published on sites like Amazon.com or Goodreads. Public reviews are especially useful for indicating the degree to which a scholarly book has become visible online and has become a topic for social engagement (Kousha et al., 2017; Zuccala et al., 2015). Both scholarly and public reviews can always be used in conjunction with other types of indicators (e.g., publisher quality and/or citations); but for improving the coverage of books in commercial citation indexes, preference is given to the scholarly review (Gorraiz et al., 2014).

*Library holding counts*

Until now, the most promising of all book-based indicators is the library holding count (Torres-Salinas & Moed, 2009; White et al., 2010; White & Zuccala, 2018 in press), which White et al. (2010) refer to as the *lib-citation* (p. 1083). A theory of lib-citation rationalizes that a holding count or set of holding counts in library catalogs might be used to indicate and calculate the perceived cultural benefit of a book or books. The advantage of this measure is that it "can make an author in the humanities look good", particularly if (s)he is not well represented in other types of databases, like the Web of Science, Google Scholar, or Scopus. White (2010) further explains that

> *on the book front, libcitations reflect what librarians*
> *know about the prestige of publishers, the opinions of reviewers,*
> *and the reputations of authors. The latter may be colored by,*
> *for example, authors' academic affiliations, previous sales, prizes,*
> *awards, distinguished appointments, mass media coverage,*
> *Web presence, and citedness. All of these are signals of what*
> *readers are likely to want, and librarians must be attuned*
> *to them* (p. 1084).

When working with this indicator, there can be at least two different methodological approaches. Torres-Salinas & Moed (2009) focus on library holding counts at the publisher level, while White et al. (2010) propose developing it at the book level. At the book level, holding counts have much more power to discriminate between books than citation counts. Records for books tend to be more plentiful in library catalogs than in citation indexes, particularly in a union catalog like OCLC-WorldCat (Zuccala & Guns, 2013). Libcitation counts for individual scholars or academic departments can be field-normalized or assigned to percentiles just as citations are. By determining how many libcitations a book needs in order to reach a 90th or 50th percentile cut-point in its main Dewey class, one can observe its cultural impact, or degree of fame relative to other titles from the same class (White & Zuccala, 2018 in press). Research also points to the fact that libcitations and citations can be statistically correlated, but one is likely to find a weak, but significant result (Zuccala & White, 2015). Both the citation and the libcitation capture a certain amount of scholarly impact in common, but this seems to be truer when holding counts are obtained from academic libraries, rather than other types of libraries. Another study based on the Altmetric suite PlumX, now shows that out of 18 types of indicators for books, including citations, downloads, views, and social media mentions, the most informative is the library holding count (Torres-Salinas et al., 2017).

## 6. Integrating book metrics into evaluation practices

For some time, the social sciences and humanities have either been assessed partially or neglected entirely due to the lack of data available for developing promising book metrics. Acceptance of this fact grew in part because of the increasing value of journal articles (in most fields), notwithstanding the long tradition of relying on journal citation indexes for many international research evaluation procedures. Auspiciously, this did not stop some of the early bibliometric monitors from examining the role of books in book-oriented research disciplines, nor did it prevent commercial organizations like Elsevier and Clarivate Analytics (formerly

Thomson Reuters) from addressing the data gap by developing a Scopus index of books and Book Citation Index℠. Subject classifiers and indexers now have ample reasons to step to the forefront, not only to apply research to these indexes, but lead the bibliometrics community forward to an improved situation: one in which exploring books and their publishers metrically is no longer an aspiration, but an established reality.

Still, the integration of books into evaluation practices will never be left solely to commercial data providers or researchers. National policy makers are stakeholders in the evaluation game and also play a role. Research by Giménez-Toledo et al. (2016) and Williams et al. (2018) provide valuable overviews of how countries across Europe have recently been implementing policies and strategies for book-based evaluations. In the United Kingdom, most scholarly books are submitted to panels C (social sciences) and D (humanities) of the panel-based Research Excellence Framework (REF). Since the panels (as well as sub-panels) take into consideration what is most valued in these broader disciplines, a qualitative approach to evaluation is used. A different approach is taken in Spain, Denmark, and Finland where evaluation procedures for books are based on league tables or *authority* lists of publishers. Panels of experts are recruited here as well, but invited to participate in the development of such lists. The publisher lists are then used to benchmark the value of a monograph submitted to each country's respective evaluation exercise. Other countries like Flanders (Belgium) implement a point system by which a book's value is weighted (e.g., monographs receive four points while edited books receive one point). Norway uses a mixed method approach where publishers and journals are divided into two levels, whereby a level two designation is the most selective. Depending on the level, a monograph will receive either 5 points (i.e., for a book with a level 1 publisher) or 8 points (i.e., for a book with a level 2 publisher). Denmark generally follows Norway's approach; hence with a similar system, a fraction of funding each year is allotted to Danish university departments that achieve the highest points.

More often than not, these evaluation policies are designed for practical purposes. Again, it is simply impractical to assess the individual contribution, quality, influence, or impact of every monograph, at a national or international level. This issue, together with the uncertainty of applying citation analysis to books, and criticisms coming from social scientists and humanities researchers has prevented the widespread development and use of citation-based indicators. Policy makers have thus been keen to disregard the citation, including many other practices, in favor of focusing on publisher status (i.e., as per the league, or *authority* tables). As a result, certain challenges related to book metrics have yet to be addressed. Presently we are at the stage where disparities in data coverage (Torres-Salinas et al., 2017) and low correlations between citations to books and alternative indicators of their impact (Kousha et al., 2017; Zuccala et al., 2015; Zuccala & White, 2015) remain difficult to interpret. With citation indicators alone, differences per database at least show moderately significant correlations (Kousha et al., 2011).

From a research perspective, it is clear then that more work needs to be done to improve upon the subject classification of books, both in commercial and national indexes, and to ensure that record keeping is complete (e.g., indexes that include author affiliations and show how books belong to bibliographic families). Scholars who work with these indexes - i.e., the indicator constructionists - are urged to remain steadfast at uncovering, refining and emphasizing different elements related to the influence or impact of books. Their biggest challenge; however, may not necessarily be technical or data-oriented, but cultural in nature.

Citation-based indicators have long been associated with research assessment schemes directed towards the natural and exact sciences. Journal articles and their received citations accommodate research communities grounded upon previous work and rapid progress: a Kuhnian model of normal science. By contrast, books and their reviews fit within a 'social' view of scholarship. Here the standards are based on the perceptions of peers; it is the academic peer who determines the value of a work. In theoretical disciplines where books are most prominent, this community-based reflexivity, inherent to the overall reflexive nature of the social sciences and humanities, is likely to remain a primary strength (Flyvberg, 2001). When bibliometric approaches to evaluation focus on complementing this strength, and recognize also a book's broader (i.e., educational, social, literary) influence or impact, book-based scholarship will not change due to perceived faults in the evaluation system, but evolve because different aspects of the truth will become more evident.

20Garfield, E. (2006). The history and meaning of the journal impact factor, *JAMA Journal of the American Medical Association, 295(1)*, 90-95.

Gingras, Y. (2014). Criteria for evaluating indicators. In B. Cronin & C. R. Sugimoto, Eds., *Beyond Bibliometrics: Harnessing Multidimensional Indicators of Scholarly Impact* (pp. 109-125). The MIT Press: Cambridge, Massachusetts.

Giménez-Toledo, E., & Román-Román, A. (2009). Assessment of humanities and social sciences monographs through their publishers: a review and a study towards a model of evaluation. *Research Evaluation, 18(3)*, 201–213

Giménez-Toledo, E., Mañana-Rodríguez, J., and Tejada-Artigas, C.-M. (2015). Scholarly publishers indicators: prestige, specialization and peer review of scholarly book publishers. *El profesional de la información, 24(6)*, 855-860.

Giménez-Toledo, E., Manana-Rodrıguez, J., Engels, T. C. E., Ingwersen, P., Polonen, J., Sivertsen, G., Verleysen, F.T. and Zuccala, A. A. (2016). Taking scholarly books into account. Current developments in five European countries. *Scientometrics, 107(2)*, 685-699.

Giménez-Toledo, E., Tejada-Artigas, C., & Mañana-Rodríguez, J. (2012). Evaluation of scientific books' publishers in social sciences and humanities: Results of a survey. *Research Evaluation, 22(1),* 64 -77.

Glenn, N. (1978). On the misuse of book reviews. *Contemporary Sociology, 7(3)*, 254–255.

Glänzel, W. & Moed, H. (2002). Journal impact measures in bibliometric research. *Scientometrics, 53(2)*, 171-193.

Glänzel, W. & Schubert, A. (2003). A new classification scheme of science fields and subfields designed for scientometric evaluation purposes. *Scientometrics, 56(3)*, 357-367.

Glänzel, W., Thijs, B., & Chi, P-S. (2016). The challenges to expand bibliometric studies from periodical literature to monographic literature with a new data source: the Book Citation Index. *Scientometrics, 109(21),* 65–2179.

González-Pereira B., Guerrero-Boteb, V.P., & Moya-Anegónc, F. (2010). A new approach to the metric of journals' scientific prestige: the SJR indicator. *Journal of Informetrics, 4(3)*, 379-391.

Kousha, K., & Thelwall, M. (2016b). An automatic method for assessing the teaching impact of books from online academic syllabi. *Journal of the Association for Information Science and Technology, 67*(12), 2993-3007.

Kousha, K., & Thelwall, M. (2017). Are Wikipedia citations important evidence of the impact of scholarly articles and books? *Journal of the Association for Information Science and Technology, 68*(3), 762-779.

Kousha, K., Thelwall, M., and Abdoli, S. (2017). Goodreads reviews to assess the wider impacts of books. *Journal of the Association for Information Science and Technology*, *68(8),* 2004-2016.

Kors, A.C. (2016). *Epicureans and Atheists in France, 1650–1729*. Cambridge University Press.

Larivière, V. & Macaluso, B. (2011). Improving the coverage of social science and humanities researchers' output: The case of the Érudit journal platform. *Journal of the American Society for Information Science and Technology, 62(12),* 2437-2442.

Lewison, G. (2001). Evaluation of books as research output in the history of medicine. *Research Evaluation, 10(2)*, 89-95.

Leydesdorff, L. & Felt, U. (2012). "Books" and "book chapters" in the book citation index (BKCI) and science citation index (SCI, SoSCI, A&HCI). *Proceedings of the American Society for Information Science and Technology, 49(1),* 1-7. [DOI: 10.1002/meet.14504901027]

Leydesdorff, L. and Rafols, I. (2009). A global map of science based on the ISI Subject Categories. *Journal of the American Society for Information Science and Technology, 60(2)*, 348–362.

Lindholm-Romantschuk, Y. (1998). *Scholarly book reviewing in the social sciences and humanities. The flow of ideas within and among disciplines*. Westport, CT: Greenwood Press.

Linmans, A. J. M. (2010). Why with bibliometrics the Humanities does not need to be the weakest link. Indicators for research evaluation based on citations, library holdings, and productivity measures, *Scientometrics, 83(2),* 337-354.

Mann, M. (1930). *Introduction to cataloging and the classification of books* (2nd ed.), Chicago: American Library Association.